\documentclass[aps,floats]{revtex4}
\usepackage{amsmath,amssymb}
\usepackage{graphicx,epsfig}

\RequirePackage{color}
\definecolor{MyDarkGreen}{rgb}{0.02,0.60,0.06}

\begin{document}
\bibliographystyle {plain}

\def\oppropto{\mathop{\propto}} 
\def\opmin{\mathop{\min}} 
\def\opmax{\mathop{\max}} 
\def\oplim{\mathop{\lim}} 
\def\opsimeq{\mathop{\simeq}}
\def\opoverderline{\mathop{\overline}}
\def\operarrow{\mathop{\longrightarrow}}
\def\opsim{\mathop{\sim}}

\def\fig#1#2{\includegraphics[height=#1]{#2}}
\def\figx#1#2{\includegraphics[width=#1]{#2}}


\title{  Symmetry relation for multifractal spectra at random critical points  } 


 \author{C\'ecile Monthus $^{(1)}$, Bertrand Berche $^{(2)}$ and 
Christophe Chatelain $^{(2)}$ }
\affiliation{ $^{(1)}$ Institut de Physique Th\'{e}orique, CNRS and CEA Saclay,
 91191 Gif-sur-Yvette, France \\
$^{(2)}$ Statistical Physics Group, P2M Dpt, Institut Jean Lamour,
Nancy Universit\'e, UMR CNRS 7198,\\
BP 70239, F-54506 Vand\oe uvre les Nancy Cedex, France.\\ }

\begin{abstract}
Random critical points are generically characterized by multifractal properties. In the field of Anderson localization, Mirlin, Fyodorov, Mildenberger and Evers [Phys. Rev. Lett 97, 046803 (2006)] have proposed that the singularity spectrum $f(\alpha)$ of eigenfunctions satisfies the exact symmetry $f(2d-\alpha)=f(\alpha)+d-\alpha$ at any Anderson transition. In the present paper, we analyse the physical origin of this symmetry in relation with the Gallavotti-Cohen fluctuation relations of large deviation functions that are well-known in the field of non-equilibrium dynamics: the multifractal spectrum of the disordered model corresponds to the large deviation function of the rescaling exponent $\gamma=(\alpha-d)$ along a renormalization trajectory in the effective time $t=\ln L$.  We conclude that the symmetry discovered on the specific example of Anderson transitions should actually be satisfied at many other random critical points after an appropriate translation. For many-body random phase transitions, where the critical properties are usually analyzed in terms of the multifractal spectrum $H(a)$ and of the moments exponents $X(N)$ of two-point correlation function [A. Ludwig, Nucl. Phys. B330, 639 (1990)], the symmetry becomes $H( 2X(1) -a)= H( a )  + a-X(1)$, or equivalently $\Delta(N)=\Delta(1-N)$ for the anomalous parts $\Delta(N) \equiv  X(N)-NX(1)$. We present numerical tests in favor of this symmetry for the 2D random $Q-$state Potts model with various $Q$.
\end{abstract}

\maketitle
\section{ Introduction }

Among the various areas where multifractality appears 
(see for instance \cite{halsey,Pal_Vul,Stan_Mea,Aha,Meakin,
harte,duplantier_houches} and references therein), 
the case of critical points in the presence of frozen
disorder is of particular interest.
The idea that multifractality occurs at criticality
has been first established for quantum Anderson localization transitions
\cite{Weg,Cas_Pel} and has been the subject of 
very detailed studies (see the reviews \cite{janssenrevue,mirlinrevue}).
However multifractality is not limited to this type
of one-particle quantum transitions, but is expected to hold
at many random critical points whenever disorder is relevant,
in particular in disordered spin models like
random ferromagnets \cite{Ludwig,Jac_Car,Ols_You,Cha_Ber,Cha_Ber_Sh,PCBI,BCrevue},
 spin-glasses or 
random field spin systems \cite{Sourlas,Thi_Hil,Par_Sou},
as well as in disordered polymer models like 
directed polymer in random media \cite{DPmultif}
or disordered wetting models \cite{wettingmultif}.
The only exceptions to these multifractal behaviors
seem to be  the ``multiscaling'' behaviors \cite{multiscaling},
which are even stronger than multifractality, that have been found
for some critical correlation functions
in disordered quantum spin-chains governed by 
 ``Infinite disorder fixed points'' \cite{revueigloi}.
So the presence of multifractality
at criticality seems generic whenever disorder is relevant.

For Anderson localization models,
 Mirlin, Fyodorov, Mildenberger and Evers \cite{mirlin06}
have proposed that the singularity spectrum $f(\alpha)$ of critical
eigenfunctions satisfies some remarkable exact symmetry 
 at any Anderson transition 
(see more details in section \ref{secanderson} below).
In this paper, we analyse the physical origin of this symmetry
in relation with the well-know symmetry relations of large deviation
functions in non-equilibrium dynamical models. We conclude that 
this symmetry is not specific to Anderson transitions
but should be satisfied at many other random critical
points, in particular at many-body disordered phase transitions
after an appropriate translation that we describe. 
As an example of application, we consider 
the two-dimensional random Potts model
 with various $Q$ and present numerical tests of the symmetry
concerning the multifractal spectrum of the two-point
order parameter correlation functions.

The paper is organized as follows.
In section \ref{secanderson}, we recall the meaning of
the symmetry relation 
for Anderson localization models.
In section \ref{secanalogy}, we analyse the similarity with
the well-known fluctuation relations 
for large deviation functions occurring in the field of 
non-equilibrium dynamics. This is the central section of the present paper.
In section \ref{secmanybody}, we 
translate the symmetry relation for $f(\alpha)$
in terms of the multifractal spectrum $H(a)$ that characterizes
the statistics of two-point correlation function
at many-body random critical points.
In section \ref{secnumeri}, we present numerical tests in favor of
this symmetry for the two-dimensional random $Q-$state Potts model
 with various $Q$.
Section \ref{secconclusion} contains our conclusions.

\section{ Symmetry relation at Anderson transitions }
\label{secanderson}

\subsection{ Reminder on the singularity spectrum $f(\alpha)$ of critical eigenfunctions}

For Anderson localization transitions, 
the  multifractal spectrum $f(\alpha)$ of
 critical eigenfunctions 
is  defined as follows
(for more details see the reviews 
\cite{janssenrevue,mirlinrevue}):
in a sample of size $L^d$, the number ${\cal N}_L(\alpha)$
of points $\vec r$ where the weight $\vert \psi_L(\vec r)\vert^2$
scales as $L^{-\alpha}$ behaves as 
\begin{eqnarray}
{\cal N}_L(\alpha) \oppropto_{L \to \infty} L^{f(\alpha)}
\label{nlalpha}
\end{eqnarray}
The inverse participation ratios, which are the most convenient
order parameters of the transition, can be then rewritten
as an integral over $\alpha$
\begin{equation}
Y_q(L)  \equiv \int_{L^d} d^d { \vec r}  \vert \psi_L (\vec r) \vert^{2q}
\simeq \int d\alpha \ L^{f(\alpha)} 
\ L^{- q \alpha} \opsimeq_{L \to \infty} 
L^{ - \tau(q) }
\label{ipr}
\end{equation}
where the exponent $\tau(q)$ 
can be obtained via a saddle-point
calculation in $\alpha$ to obtain the Legendre
transform formula \cite{janssenrevue,mirlinrevue}
\begin{eqnarray}
 -\tau(q)  =  \opmax_{\alpha} \left[ f(\alpha) - q \alpha  \right]
\label{legendre}
\end{eqnarray}
The usual normalization condition of eigenfunctions 
\begin{equation}
\int_{L^d} dr \vert \psi_L (\vec r) \vert^{2}  = 1
\label{normapsi2}
\end{equation}
implies $Y_{q=1}(L)=1$
so that $\tau(q=1)=0$. The weight $\vert \psi_L(\vec r)\vert^2$
has thus for disorder averaged value
\begin{equation}
\overline{  \vert \psi_L (\vec r) \vert^{2} } = \frac{1}{L^d}
\label{avweight}
\end{equation}
whereas  the typical exponent $\alpha_{typ}$
governing the typical decay
\begin{equation}
\overline{  \ln \vert \psi_L (\vec r) \vert^{2} } \oppropto_{L \to \infty}
- \alpha_{typ} \ln L
\label{typweight}
\end{equation}
is determined by the maximum $f(\alpha_{typ})=d$ of the spectrum $f(\alpha)$.
These scaling behaviors, which concern  
individual eigenstates $\psi$, can be translated
for the local density of states
\begin{equation}
\rho_L(E,\vec r) = \sum_{n} \delta(E-E_n) \vert \psi_{E_n}(\vec r)\vert^2
\label{defrho}
\end{equation}
as follows: for large $L$, when the $L^d$ energy levels become dense,
the sum of Eq. \ref{defrho} scales as
\begin{equation}
\rho_L(E, \vec r) \propto L^d \vert \psi_E(\vec r)\vert^2
\label{equiv}
\end{equation}
and its moments involve the exponents $\tau(q)$ introduced in Eq. \ref{ipr}
\begin{equation}
\overline{ [\rho_L(E,\vec r)]^q } \oppropto_{L \to \infty}
\frac{1}{L^{\Delta(q)}} \ \ {\rm with \ \ } \Delta(q) =  \tau(q)-d (q-1) 
\label{rhomoments}
\end{equation}

\subsection{ Symmetry relation of $f(\alpha)$}

For any Anderson transition in the so-called 
'conventional symmetry classes' \cite{mirlinrevue},
 Mirlin, Fyodorov, Mildenberger and Evers \cite{mirlin06}
have proposed that the singularity spectrum $f(\alpha)$ of critical
eigenfunctions satisfies the remarkable exact symmetry 
\begin{eqnarray}
f(2d-\alpha)=f(\alpha)+d-\alpha
\label{symfaanderson}
\end{eqnarray}
that relates the regions $\alpha \leq d$ and $\alpha \geq d$.
(see for instance Fig. 1 of Ref. \cite{vasquez} b for a pictorial
representation of this symmetry).
In terms of the exponents 
$\tau(q)$ of Eq. \ref{ipr}, the symmetry
of Eq. \ref{symfaanderson}
with respect to the value $\alpha_s=d$ becomes 
a symmetry with respect to the value $q_s=1/2$ \cite{mirlin06}
\begin{eqnarray}
\tau(q)-\tau(1-q)=d(2q-1)
\label{symtauqanderson}
\end{eqnarray}
In terms of the exponents
$\Delta(q)$ of Eq. \ref{rhomoments},
the symmetry takes the simpler form
\begin{eqnarray}
\Delta(q)=\Delta(1-q)
\label{symdeltaqanderson}
\end{eqnarray}

As discussed in \cite{mirlin06}, 
this type of symmetry has been first derived for 
 non-linear sigma-models \cite{sigma}, which are a priori
only valid for weak disorder.
Nevertheless,  Mirlin, Fyodorov, Mildenberger and Evers 
have argued in \cite{mirlin06} that Eq. \ref{symfaanderson}
should remain exact even at strong disorder as a consequence  
of universality of critical properties.
Eq. \ref{symfaanderson} has been checked 
on the exact expansion in $d+\epsilon$ up to 4-loop order
\cite{Wegner_epsilon}, and numerically 
for various types of Anderson transitions, in particular in the Power-Law
random banded matrices \cite{mirlin06},
in the symplectic 2D Anderson transition \cite{milden07},
in the quantum Hall transition \cite{qHall}, and
in the 3D Anderson transition \cite{vasquez}
 (the only exception being, 
to our knowledge, the spin quantum Hall transition
\cite{surface} that belongs to the Bogoliubov-de Gennes symmetry class C
of the symmetry classification \cite{mirlinrevue})
This symmetry has even been measured in recent experiments \cite{experiment}.

In conclusion, the symmetry of Eq. \ref{symfaanderson}
seems very well satisfied at most of Anderson transitions
where it has been studied.
 However its physical origin
has remained unclear.
In particular, an important issue
is whether this symmetry is specific to 
critical theories of Anderson transitions,
or whether  it could be
satisfied at other random critical points. 
These questions have been the motivations of the present work.

\section{ Physical origin of the symmetry relation  }

\label{secanalogy}

The multifractal formalism can be seen
as a theory of large deviations in the parameter $\ln L$
(see \cite{touchette}
for a recent review on large deviations in statistical physics).
It is thus interesting to discuss in this section
the similarity that exists 
between the Mirlin-Fyodorov-Mildenberger-Evers symmetry of Anderson
transitions described in the previous section, and the 
well-known symmetry relations
 of large deviation functions that occur in the field
of non-equilibrium dynamics.

\subsection{ Fluctuations relations
of large deviations functions in non-equilibrium dynamical models }

\label{subsec:NonEq}

It is clearly impossible to summarize here all the recent developments
concerning the various 'fluctuation relations' 
that have been established 
in the field of non-equilibrium dynamics recently,
and we refer the interested reader to some recent reviews 
\cite{derrida,harris_Schu,kurchan,searles,zia,maes,chetrite} and references therein.
Here we will simply recall some basic definitions
that will be useful for our present purposes.
The observables $Y_t$ which are expected
 to become extensive in time in the large time limit, 
usually satisfy some large deviation principle: 
the probability to have a given time-averaged value $Y_t/t=j$
behaves at large $t$ as
\begin{eqnarray}
Prob\left( \frac{Y_t}{t} =j  \right)  \oppropto_{t \to +\infty} 
e^{ t G(j) }
\label{noneqGj}
\end{eqnarray}
where $G(j) \leq 0$ is called
 the large-deviation function.
The typical value $j_{typ}$ corresponds to the point
where it vanishes $G(j_{typ})=0$.
It has been found that in many cases,  the large deviation function $G(j)$
satisfies some symmetry relation of the form
\begin{eqnarray}
G(j)=G(-j)+ K j
\label{symGj}
\end{eqnarray}
where the constant $K$ may contain the physical parameters
for the problem at hand (like the reservoirs fugacities,
the temperature, the external applied field, etc).
Equivalently, the generating function of the variable $Y_t$
behaves for large time $t$ as
\begin{eqnarray}
\langle  e^{\lambda Y_t} \rangle  = \int dj e^{ t (\lambda j + G(j) )}
\oppropto_{t \to \infty} e^{t \mu(\lambda)}
\label{noneqmu}
\end{eqnarray}
where $\mu(\lambda)$ is the Legendre transform of $G(j)$
\begin{eqnarray}
 \mu(\lambda) = \opmax_{j} \left[ \lambda j + G(j) \right]
\label{mulambdalegendre}
\end{eqnarray}
Its series expansion in $\lambda$ 
\begin{eqnarray}
\mu(\lambda) \equiv  \oplim_{t \to +\infty}   \frac{ \ln  \langle  e^{\lambda Y_t}    \rangle }{  t} 
= \lambda  j_{typ}  +\frac{\lambda^2  }{2} \sigma +o(\lambda^2)
\label{generatingcumulants}
\end{eqnarray}
 yields the successive cumulants of $Y_t$
\begin{eqnarray}
j_{typ} && =  \oplim_{t \to +\infty}  \frac{  \langle   Y_t  \rangle }{  t }  \nonumber \\
\sigma && =\oplim_{t \to +\infty}  \frac{ \langle   Y_t^2 \rangle  - \langle   Y_t\rangle ^2 }{  t }  
\label{firstcumulants}
\end{eqnarray}
The symmetry of Eq. \ref{symGj} reads for the generating function $\mu(\lambda)$
\begin{eqnarray}
\mu(\lambda)= \mu(-K-\lambda)
\label{symmu}
\end{eqnarray}

The symmetry of Eq. \ref{symGj} or Eq. \ref{symmu}
is usually called a Gallavotti-Cohen fluctuation relation.
The full domain of validity of this type of symmetry
in the field of non-equilibrium dynamics
is not easy to state, since it has been derived
in various contexts with different assumptions
 on the dynamics, which can be either  stochastic
or deterministic but sufficiently chaotic 
(see the various presentations in the reviews 
\cite{derrida,harris_Schu,kurchan,searles,zia,maes,chetrite}).
Let us remind here the most important ideas on the simplest case:
for a stochastic dynamics defined by some Markov chain
with transition probabilities $k(C \to C')$ between configurations 
\begin{eqnarray}
P_{t+1}(C) = \sum_{C'}  P_t(C') k(C' \to C)
\label{master}
\end{eqnarray}
the symmetry relation of large deviation has usually for
origin some 'generalized detailed balance relation' of the form
\begin{eqnarray}
 k_{y}(C \to C') = k_{-y}(C' \to C) e^{ K y}
\label{generalizeddetailedbalance}
\end{eqnarray}
where $y$ denotes the increase of the dynamical quantity
$Y_t$ for the jump $C \to C'$, and $(-y)$ denotes
 the increase of the dynamical quantity
$Y_t$ for the jump $C' \to C$.
 Eq. \ref{generalizeddetailedbalance} implies that 
a given dynamical trajectory $Traj= \{C_0,C_1,..C_t\}$ characterized by the value 
$Y_t(Traj)=y_1+y_2+..+y_t$, and the reversed trajectory $(-Traj)= \{C_t,C_{t-1},..C_0\}$ 
characterized by the opposite value $Y_t(-Traj)=-Y_t(Traj)$
have probabilities related by the simple relation
\begin{eqnarray}
\frac{ P(Traj) }{P(-Traj)} =  \frac{k_{y_1}(C_0 \to C_1)k_{y_2}(C_1 \to C_2) ... k_{y_t}(C_{t-1} \to C_t)}
{k_{-y_t}(C_t \to C_{t-1})...k_{-y_2}(C_2 \to C_1)  k_{-y_1}(C_1 \to C_0)}  = e^{ K Y_t(Traj) }
\label{ratiotrajdynamics}
\end{eqnarray}

The generating function of 
Eq. \ref{noneqmu} reads
\begin{eqnarray}
\langle  e^{\lambda Y_t} \rangle   \equiv \sum_{Traj} {P}(Traj) e^{\lambda Y_t(Traj)}
= \sum_{Traj} P(-Traj) e^{(K+\lambda) Y_t(Traj)}
\label{genetraj}
\end{eqnarray}
 Using the one-to-one change of variable 
$Traj'=-Traj$ and the antisymmetry relation $Y_t(-Traj)=-Y_t(Traj)$,
one obtains
\begin{eqnarray}
\langle  e^{\lambda Y_t} \rangle  
= \sum_{Traj'} P(Traj') e^{- (K+\lambda) Y_t(Traj' )}
= \langle  e^{-(K+\lambda) Y_t} \rangle  
\label{genetrajinter}
\end{eqnarray}
which corresponds to the symmetry relation of Eq. \ref{symmu}
for the generating function introduced in Eq. \ref{noneqmu}.

To derive a symmetry relation for a large deviation function of some 
dynamical observable $Y_t$,
it is sufficient to justify 
some 'generalized detailed balance relation' of the form
of Eq. \ref{generalizeddetailedbalance} for the elementary
transition probabilities. We refer to the reviews \cite{derrida,harris_Schu,kurchan,searles,zia,maes,chetrite}
for the description of various physical situations where this can be done
(system in contact with two reservoirs at different temperatures,
system in an external applied field,etc).
However, as explained in \cite{Leb_spo}, for any stochastic process
 there is always a quantity $B_t$ for which it works by construction:
it is the quantity $B_t$ whose increment $b(C,C')$ 
is { \it defined } by Eq. \ref{generalizeddetailedbalance} with $K=1$,
i.e. it corresponds to the logarithm of the ratio of the two transition rates
\cite{Leb_spo}
\begin{eqnarray}
b(C,C') = \ln  \left( \frac{k(C \to C')}{k(C' \to C)} \right)
\label{exLebSpo}
\end{eqnarray}
The functional $B_t$ of the trajectory $Traj= \{C_0,C_1,..C_t\}$
for which the fluctuation relation holds by construction then reads \cite{Leb_spo}
\begin{eqnarray}
B_t(Traj= \{C_0,C_1,..C_t\}) = \sum_{ m=1}^t \ln  \left( \frac{k(C_{m-1} \to C_m)}{k(C_m \to C_{m-1})} \right)
\label{exLebSpototal}
\end{eqnarray}
We stop here this brief reminder on large deviation functions
occurring in non-equilibrium dynamics,
and refer the interested reader to \cite{Leb_spo} and to the reviews \cite{derrida,harris_Schu,kurchan,searles,zia,maes,chetrite}
for the physical interpretation in terms of irreversibility and entropy production.
We now analyse the analogy with multifractal spectra at random critical points.

\subsection{ Dictionary of the analogy}

To make the formal analogy between Eq. \ref{symGj}
and Eq. \ref{symfaanderson} complete,
it is convenient to replace the singularity exponent
$\alpha$ by its distance to $d$
 which represents the 'homogeneous' value under rescaling
\begin{eqnarray}
\gamma \equiv \alpha -d 
\label{defgamma}
\end{eqnarray}
Then the negative function 
\begin{eqnarray}
g(\gamma) \equiv f(\alpha= d+\gamma)-d  
\label{defggamma}
\end{eqnarray}
satisfies (Eq. \ref{symfaanderson})
\begin{eqnarray}
g(\gamma)=g(-\gamma)+\gamma
\label{symfgg}
\end{eqnarray}
This equation is now exactly similar to Eq. \ref{symGj} with $K=1$.
The relation between the generating function
$\mu(\lambda)$ of Eq. \ref{mulambdalegendre}
and the exponents $\Delta(q)$ of Eq. \ref{rhomoments}
reads 
\begin{eqnarray}
 \mu(\lambda) && = \opmax_{\gamma} \left[ \lambda \gamma + g(\gamma) \right] 
\nonumber \\
&& = \opmax_{\alpha} \left[ \lambda (\alpha-d) + f(\alpha)-d   \right]
= - \tau(-\lambda) -d \lambda -d \nonumber \\
&&
= -\Delta(-\lambda)
\label{relationmudelta}
\end{eqnarray}
so that the symmetry relation of Eq. \ref{symmu} with $K=1$
is equivalent to Eq. \ref{symdeltaqanderson}.

Since the physical situations of a non-equilibrium dynamical system and 
of a static disordered system at criticality
are a priori completely different,
this analogy may seem completely formal at first sight.
It is however very suggestive:
the large time $t$ of the dynamical system
corresponds to the logarithm  of the large linear scale $L$
for the random critical system
\begin{eqnarray}
t=\ln L
\label{timeRG}
\end{eqnarray}
The 'effective dynamics' for the random system thus corresponds
to renormalization process in scale $\ln L$.

\subsection{ Stability of the multifractal spectrum upon renormalization }

In critical phenomena, it is well known that critical properties
are stable under coarse-graining. This explains their
universal character (independence with respect to microscopic details)
and why renormalization is an appropriate framework.
Similarly for random critical points, the multifractal spectrum 
is expected to be stable under coarse-graining,
i.e. the formulations we have written above
in terms of the microscopic observables
(like the eigenfunction weight $\psi^2(r)$ in Anderson localization models)
can be reformulated in terms of coarse-grained observables as follows
\cite{janssenrevue}:
if the system of volume $L^d$ is decomposed into boxes of volume $L_b^d$
one introduces the weight $w_L^{L_b}(r)$ of the box
of volume $L_b^d$ around the point $r$ 
\begin{eqnarray}
w_L^{L_b}(r) \equiv  \int_{L_b^d} d^d r' \psi_L^2(r+r')
\label{weightbox}
\end{eqnarray}
that generalizes the microscopic weight $\psi_L^2(r)$ corresponding to $L_b=1$.
Then all previous multifractal notions apply to the box weights
if one replaces $L$ by the ratio $L/L_b$ of the two length scales.
Equation \ref{nlalpha} becomes
\begin{eqnarray}
Prob \left[ w_L^{L_b}(r) \sim \left(\frac{L_b}{L}\right)^{\alpha} \right]
\sim   \left(\frac{L}{L_b}\right)^{f(\alpha)-d}
\label{probaweightbox}
\end{eqnarray}
and equivalently, their moments behave as
\begin{eqnarray}
\overline{ \left[ w_L^{L_b}(r) \right]^q }
\sim \int d\alpha    \left(\frac{L}{L_b}\right)^{f(\alpha)-d-q \alpha}
\sim \left(\frac{L}{L_b}\right)^{-d-\tau(q)}
\label{momentsweightbox}
\end{eqnarray}
This formulation with coarse-grained boxes is also sometimes used numerically,
in particular to measure correctly
the scaling behaviors of negative moments $q<0$ \cite{mirlin06,vasquez}.

\subsection{ Analysis of the renormalization process }

To go from the microscopic
 scale $l=1$ to the macroscopic scale $l=L$
of the whole system, we may introduce intermediate scales $l_m$
regularly placed on a logarithmic scale.
For definiteness, let us consider a discrete system with $L=2^M$ 
and introduce the intermediate scales 
\begin{eqnarray}
l_m=2^m \ \ {\rm with } \ \ m=0,1,..,M
\label{deflm}
\end{eqnarray}

We are interested into the flow of
microscopic weight $\psi^2(r)$ upon renormalization.
Up to now, we had always assumed the usual normalization of Eq. \ref{normapsi2}
 on the full sample of volume $L^d$,
but here to analyse the coarse-graining process using samples of various sizes,
it is more convenient to work with fields $\psi^2(r)$
 free of any global normalization constraint.
To characterize the microscopic weight $\psi^2(r)$
within the box of volume $l_m^d$ surrounding the point $r$,
we introduce the variables
\begin{eqnarray}
W_m \equiv \frac{l_m^d \psi^2(r)}{\int_{l_m^d} d^d r' \psi^2(r+r') } 
\label{defWm}
\end{eqnarray}
If $\psi^2(r+r')$ were constant within the box of volume $l_m^d$,
one would have $W_m=1$. 
The lowest scale $l_0=1$ is characterized by $W_0=\psi^2(r)/\psi^2(r)=1$.
We are interested into the renormalization process 
\begin{eqnarray}
W_0 =1 \to W_1 \to W_2 \to ...  \to W_M \equiv 
\frac{L^d \psi^2(r)}{\int_{{L}^d} d^d r' \psi^2(r+r') }
\label{RGprocess}
\end{eqnarray}
Since the box of size $l_m^d$ contains the box of size $l_{m-1}^d$,
it is convenient to introduce the positive exponent $\hat \alpha_m \geq 0$
\begin{eqnarray}
2^{ \hat \alpha_m} \equiv \frac{\int_{l_m^d} d^d r' \psi^2(r+r')}
{\int_{l_{m-1}^d} d^d r' \psi^2(r+r')}
\label{defalpham}
\end{eqnarray}
to characterize the coarse-graining step from $l_{m-1}$ to $l_m$.
The ratio of two successive rescaled variables $W_m$ of Eq. \ref{defWm}
then reads
\begin{eqnarray}
\frac{W_m}{W_{m-1}}= 
\frac{2^d \int_{l_{m-1}^d} d^d r' \psi^2(r+r')}
{\int_{l_m^d} d^d r' \psi^2(r+r') } 
= 2^{- \hat \gamma_m }
\label{ratioWm}
\end{eqnarray}
in terms of the exponent
\begin{eqnarray}
 \hat \gamma_m \equiv \hat \alpha_m -d
\label{defgammam}
\end{eqnarray}

The variable $W_m$ can be then written as
\begin{eqnarray}
 W_m = 2^{ - \Gamma_m }
\label{Wmsumgammam}
\end{eqnarray}
in terms of the accumulated value
\begin{eqnarray}
\Gamma_m \equiv \sum_{n=1}^m \hat \gamma_n
\label{sumgamma}
\end{eqnarray}
along the RG trajectory.
The multifractal definition for the weight 
\begin{eqnarray}
Prob \left( \frac{\psi^2(r)}{\int_{{L}^d} d^d r' \psi^2(r+r')} 
\sim \frac{1}{L^{d+\gamma}} \right) \oppropto_{L \to +\infty}
L^{g(\gamma)} d\gamma
\label{multifgamma}
\end{eqnarray}
in terms of the function 
$g(\gamma)$ introduced in Eq. \ref{defggamma},
corresponds to the following large deviation behavior
for the variable $\Gamma_M$ as $M= \ln L/\ln2$ becomes large
\begin{eqnarray}
Prob \left( \frac{\Gamma_M}{M} = \gamma \right) 
\oppropto_{M \to +\infty} 2^{ M g(\gamma) }
\label{largedevgamma}
\end{eqnarray}
Equivalently, the definition of the exponents $\Delta(q)$ 
from the weight moments at Anderson transition
\begin{eqnarray}
\overline{ \left( \frac{L^d \psi^2(r)}
{\int_{{L}^d} d^d r' \psi^2(r+r')}  \right)^q }
\oppropto_{L \to +\infty} L^{- \Delta(q) }
\label{multifdeltaq}
\end{eqnarray}
corresponds, in the large deviation theory of the variable
$\Gamma_M$, to the following statement for its generating function
\begin{eqnarray}
\langle  2^{ -q \Gamma_M }  \rangle  \oppropto_{M \to +\infty} 2^{- M \Delta(q) }
\label{generatingdeltaq}
\end{eqnarray}

The symmetry property of Eq. \ref{symfgg} for $g(\gamma)$
means that  the probability to obtain a value $\Gamma_M= M \gamma$
and the probability to obtain the opposite value 
$\Gamma_M= - M \gamma $ are related via
\begin{eqnarray}
\frac{ Prob( \Gamma_M=\gamma M ) } { Prob (\Gamma_M=- \gamma M) } \oppropto_{M \to +\infty}  2^{ M \left[
g(\gamma) - g(-\gamma) \right] }
= 2^{M \gamma} = 2^{ \Gamma_M }
\label{ratioplusmoins}
\end{eqnarray}

We now come to our central assumption. {\em It seems physically natural to 
consider that the renormalization
process of Eq. \ref{RGprocess} represents some Markov chain,
i.e. one expects that the probability to see $W_m$ at scale $l_m$
will depend on $W_{m-1}$ 
but will not depend on the previous values $W_n$ with $n<m-1$
that describe the finer structure at smaller scales inside the box of volume $l_{m-1}^d$.}
The analogy with the analysis of non-equilibrium
stochastic dynamics recalled in section \ref{subsec:NonEq}
then suggests that the physical origin of Eq. \ref{ratioplusmoins}
is that the increment $\hat \gamma $ 
of the considered functional $\Gamma_M$
exactly measures the ratio of two opposite transition probabilities
as in Eq. \ref{exLebSpo}, i.e. with our present notations
\begin{eqnarray}
 2^{\hat \gamma}= \frac{ k_{\hat \gamma }(W \to W') }
{ k_{- \hat \gamma }(W' \to W) }  
\label{kwoppsite}
\end{eqnarray}
And indeed, using Eq. \ref{ratioWm}, if one writes an elementary
transition probability as 
\begin{eqnarray}
 k_{\hat \gamma }(W \to W') = \delta(W'-  2^{- \hat \gamma} W ) 
\label{kdelta}
\end{eqnarray}
one obtains, using the properties of the $\delta$-function,
that Eq. \ref{kwoppsite} is satisfied
\begin{eqnarray}
 k_{\hat \gamma }(W \to W') 
=  2^{\hat \gamma} \delta(W-  2^{ \hat \gamma} W' ) 
= 2^{\hat \gamma}  k_{- \hat \gamma }(W' \to W)
\label{kdeltacalcul}
\end{eqnarray}
{\em This interpretation suggests that the symmetry relation of
Eq. \ref{ratioplusmoins} has an origin which is completely
independent of the physical problem under consideration,
so that it should be valid not only for Anderson transitions,
but also for other random critical points.}
We should stress that for each random critical point, 
the large deviation function
$g(\gamma)$ is of course non-trivial and depends on the critical model,
but the symmetry of Eq. \ref{symfgg} seems generic and model-independent.

The physical meaning of Eq. \ref{kwoppsite} is that the variable $\hat \gamma$
characterizes the irreversibility of the RG flow.
In a pure homogeneous system, the stability of the fixed point
corresponds to $ \gamma=0$, so that the forward RG and the backward RG are equivalent.
 In the presence of quenched disorder however, the exponent $\gamma$
becomes distributed, and its typical value $\gamma_{typ}$ corresponding to $g(\gamma_{typ})=0$
is strictly positive
\begin{eqnarray}
 \gamma_{typ} >0
\label{gammatyp}
\end{eqnarray}
This introduces some asymmetry between the forward RG flow and the backward RG flow. 
The strong multifractality limit where $ \gamma_{typ} $ is far from zero corresponds
to the far-from-equilibrium limit for dynamical systems,
whereas the weak multifractality limit where $ \gamma_{typ} $ is near zero
corresponds to the close-to-equilibrium limit for dynamical systems.
Of course for a given random critical point, the disorder strength which has already been tuned
to reach criticality cannot be tuned again to reach the regime of weak multifractality,
in contrast with dynamical systems where one may always imagine to apply a weak external field
to remain close-to-equilibrium. However, for a given type of random transition,
there exists usually a parameter that allows to interpolate between weak multifractality
and strong multifractality. For the usual Anderson transitions for instance, it is the space dimension $d$
that leads to  weak multifractality for $d=2+\epsilon$ and to strong multifractality for large $d$ \cite{mirlinrevue}.
Similarly for two-dimensional random Potts models that will be considered in section \ref{secnumeri},
 it is the number $Q$ of possible states of an individual spin
that allows to interpolate weak multifractality for $Q=2+\epsilon$ and strong multifractality for large $Q$.

\section{ Symmetry relation for many-body random critical points }
\label{secmanybody}

In the previous section, we have interpreted the
symmetry of the multifractal spectrum
discovered on the specific example of Anderson transitions
as a Gallavotti-Cohen symmetry relation for a renormalization procedure.
 This suggests that this symmetry 
 should be satisfied at other random critical points after an appropriate translation.
In the remaining of this paper, we thus consider the case
of many-body random critical points, like disordered spin models,
where the multifractal properties are usually defined in terms of the statistics
of the two-point correlation.

\subsection{ Multifractal spectrum $H(a)$ of the two-point correlation function }

Following the notations of Ludwig \cite{Ludwig},
let us call $\phi(r)$ the local order parameter,
and $X(N)$ the scaling dimensions of its disorder-averaged 
moments $\overline{\phi^N(r)}$.
Then the two-point correlation function
\begin{eqnarray}
C(r,r+R) \equiv \langle  \phi(r) \phi(r+R) \rangle 
\label{corre}
\end{eqnarray}
will present the following scaling behaviors
for its disorder-averaged moments
\begin{eqnarray}
\overline{C^N(r,r+R) } = \int dC C^N P_R(C) \oppropto_{R \to +\infty}
 \frac{1}{R^{2 X(N)}}
\label{momentscorre}
\end{eqnarray}
Equivalently, the probability distribution $P_R(C)$
is described by the multifractal spectrum $H(a) \geq 0$ such that
\begin{eqnarray}
dC P_R \left( C \sim \frac{1}{R^{2a}} \right) 
   \oppropto_{R \to +\infty} R^{- 2 H(a)} da
\label{probacorre}
\end{eqnarray}
The saddle-point calculation of the integral in Eq. \ref{momentscorre}
yields that 
$X(N)$ is the Legendre transform of $H(a)$
\begin{eqnarray}
-2 X(N) = \opmax_{a} \left[ -2 N a - 2 H(a) \right]
\label{legendreXH}
\end{eqnarray}
The minimum $a_{typ}$ of $H(a)$ where $H(a_{typ})=0$ governs
 the typical correlation
\begin{eqnarray}
\overline{\ln C(r,r+R) } = - 2 a_{typ} \ln R
\label{typcorre}
\end{eqnarray}
whereas the averaged correlation is governed by $X(1)$
\begin{eqnarray}
\overline{C(r,r+R) }  \oppropto_{R \to +\infty} \frac{1}{R^{2 X(1)}}
\label{avcorre}
\end{eqnarray}

\subsection{ Translation of the symmetry of $f(\alpha)$
into a symmetry relation for $H(a)$ }
\label{sectranslation}

To translate the symmetry relation that holds 
for the singularity spectrum $f(\alpha)$ in Anderson localisation models
to the many-body transitions, we should first rephrase
the statements concerning two-point functions within an infinite system
into statements concerning one-point functions within a finite-size system.
In a system of size $L^d$, the local order parameter $\phi_L(r)$
is characterized by the scaling dimensions $X(N)$ of Eq. \ref{momentscorre}
that govern the moments
\begin{eqnarray}
\overline{\phi^N_L(r) }  \oppropto_{L \to +\infty}
 \frac{1}{L^{ X(N)}}
\label{momentsonepoint}
\end{eqnarray}
or by the multifractal spectrum $H(a)$ of Eq. \ref{probacorre}
\begin{eqnarray}
d\phi P_R \left( \phi \sim \frac{1}{L^{a}} \right) 
   \oppropto_{L \to +\infty} L^{-  H(a)} da
\label{probaonepoint}
\end{eqnarray}
Now to make the link with the Anderson localization framework
or more generally with the usual multifractal formalism,
which is defined
in terms of a normalized probability measure \cite{halsey},
it is convenient to construct a probability measure
from the non-normalized observables one is interested in
 \cite{Dup_Lud,Pook,wettingmultif}.
We thus introduce the normalized local order parameter
\begin{eqnarray}
w_L(r) \equiv \frac{\phi_L(r)}{ \int_{L^d} d^d r' \phi_L(r) }
\label{weightonepoint}
\end{eqnarray}
where the denominator scaling is
 governed by the exponent $X(1)$ of Eq. 
\ref{momentsonepoint} as a consequence of the equivalence between
spatial-averaging and disorder-averaging at the scaling level
\begin{eqnarray}
 \int_{L^d} d^d r' \phi_L(r) \sim L^{d-X(1)}
\label{avonepoint}
\end{eqnarray}

The $f(\alpha)$ analogous to the Anderson localization
definition of Eq. \ref{nlalpha} is that the probability to have 
 $w_L(r) \sim 1/L^{\alpha}$ behaves as
\begin{eqnarray}
dw P_R \left( w_L(r) \sim \frac{1}{L^{\alpha}} \right) \sim L^{f(\alpha)-d} 
d\alpha
\label{probaweight}
\end{eqnarray}
If $w_L(r) \sim 1/L^{\alpha}$,
the one-point function decays as $\phi_L(r)\sim 1/L^a$
with (see Eqs \ref{weightonepoint} and \ref{avonepoint})
\begin{eqnarray}
a= \alpha - d+X(1)
\label{aetalpha}
\end{eqnarray}
and thus the relation between the multifractal spectra $f(\alpha)$ and $H(a)$
reads (Eqs \ref{probaweight} and \ref{probaonepoint})
\begin{eqnarray}
f(\alpha)-d = -H\left( a=\alpha - d+X(1) \right)
\label{Haetfalpha}
\end{eqnarray}
Equivalently, the relation between their respective 
Legendre transforms $\tau(q)$ and $X(N)$ reads
\begin{eqnarray}
\tau(N) = X(N)-NX(1)+d(N-1)
\label{tauetX}
\end{eqnarray}
i.e. the exponents $\Delta(N)$ introduced in Eq. \ref{rhomoments}
\begin{equation}
\Delta(N) =  \tau(N)-d (N-1) =  X(N)-NX(1)
\label{deltaetX}
\end{equation}
directly measure the non-linearity of $X(N)$.

The symmetry relation of Eq. \ref{symfaanderson} translates for $H(a)$ into
\begin{eqnarray}
  H\left( 2X(1) -a  \right)
= H\left( a \right)  + a-X(1)
\label{symHapotts}
\end{eqnarray}
i.e. a symmetry with respect to $a_s=X_1$.
In terms of the Legendre transform $X(N)$, the symmetry becomes
\begin{equation}
X(N)-X(1-N)= (2N-1) X(1)
\label{SymXN}
\end{equation}
which is analogous to Eq. \ref{symtauqanderson}.
Finally, in terms of $\Delta(N)$ of Eq. \ref{deltaetX},
the symmetry is given by Eq. \ref{symdeltaqanderson} again
\begin{eqnarray}
\Delta(N)=\Delta(1-N)
\label{symdeltaN}
\end{eqnarray}

\section{Tests of the symmetry for the 2D random Potts model}

\label{secnumeri}

The  random many-body transition where multifractal properties have been the
most studied, is the disordered two-dimensional ferromagnetic Potts model 
 \cite{Ludwig,Lewis,Jac_Car,Ols_You,Cha_Ber,Cha_Ber_Sh,PCBI,BCrevue}:
it is a lattice spin model defined by the
Hamiltonian
	\begin{equation}
	-\beta H=\sum_{(i,j)} J_{ij}\delta_{\sigma_i,\sigma_j}
	\label{defPotts}
	\end{equation}
where the spin variables can take $Q$ different values 
$\sigma_i\in \{0,\ldots,Q-1\}$. The sum extends 
over nearest neighbors on the lattice. The exchange couplings $J_{ij}$ are
quenched variables. In the following, we have considered
 two different types of disorder realizations:
 i)  self-dual random 
bond realizations, i.e. two different non zero `ferromagnetic' 
values of the nearest neighbour
interaction distributed with the same probability $\frac 12$, 
and ii) dilute bond,
i.e. non zero values distributed over a fraction of the links only. 
The two types of
disorder realizations belong to the same universality class.
The phase transition is of second order for any value of $Q$
in the disordered case (whereas in the pure case, the transition is 
of second order for $Q\le4$
and of first order for $Q>4$).

The moments of the spin-spin correlation function
are expected to decay algebraically
	\begin{equation}
	\overline{\langle\delta_{\sigma_i,\sigma_j}\rangle^N}\sim |\vec r_i-\vec r_j|^{-2 X(N)}
	\end{equation}
where we keep the prescription that  
$\overline{\ldots\vphantom{A}}$ denotes the average over disorder while $\langle\ldots \rangle$ is the thermal average.
The exponents $X(N)$ are the exponents
that have been
 introduced in the previous section in Eq. \ref{momentscorre} in the general 
case.

\subsection{ Numerical tests of the symmetry
of the multifractal spectrum for various $2 < Q \leq 8$}

We now present numerical data obtained using the connectivity
transfer matrix introduced by Bl\"ote and Nightingale~\cite{BloteNightingale}.
In this formulation,  even non-integer values of $Q$
can be considered because the number $Q$ of states enters
only as a parameter. Furthermore 
large strips can be considered because the dimension
of the transfer matrix is smaller in the connectivity space than in the spin
space where it grows as $Q^L\times Q^L$ for strip width $L$. 
 Finally the spin-spin correlation functions can be
expressed in terms of connectivity and thus be computed very efficiently.
In the following, the data have been averaged
 over 80,000 disorder realizations.
Further details about the numerical procedure can be found in~\cite{Cha_Ber}.

\begin{figure}[!th]
\epsfxsize=16cm\epsfbox{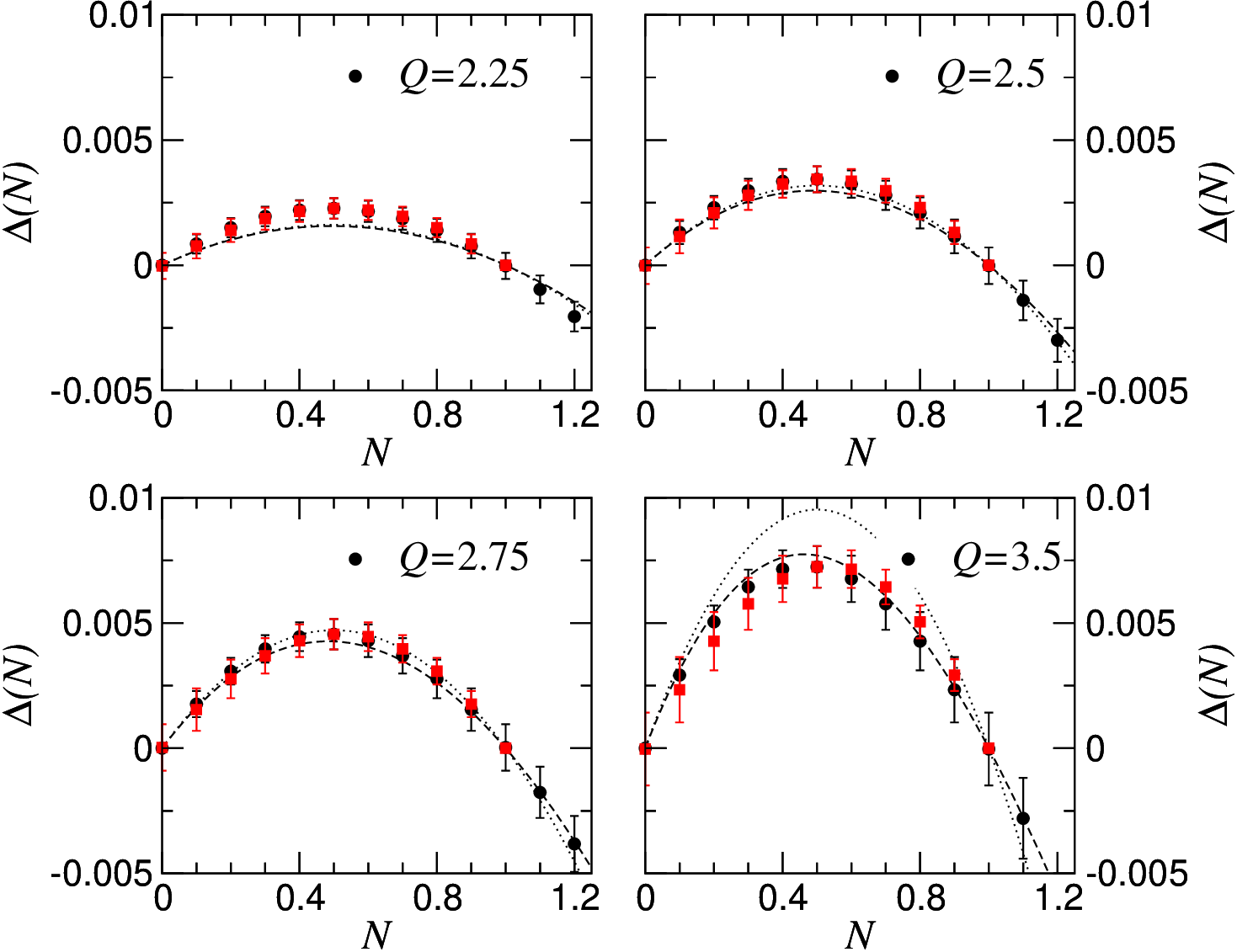}
\caption{Numerical test of the symmetry \ref{symdeltaN} for the
quantity $\Delta(N)$, calculated from the exponents $X(N)$ associated
to the decay of the moments $\overline{\langle \sigma_i\sigma_{i+u}\rangle^N}$
for the self-dual random bond Potts model for $Q \le 3.5$. The red symbols
corresponds to the data after the operation $N\rightarrow 1-N$ while the
black ones are the original data. 
The dashed line are the perturbative expansions:
the first order of Eq. \ref{ExpLudwig} is shown with short dashed lines,
whereas the second order of Eq. \ref{resLewis} is shown long dashed lines. }
\label{fig3}
\end{figure}

The system consists in an infinitely long strip of finite 
width $L$ with periodic
boundary conditions in the transverse direction, 
i.e. it is asymptotically one-dimensional. 
Therefore the correlations
decay exponentially rather than algebraically at the critical point
and as a consequence of conformal invariance, the average correlation functions
and its moments involve the critical exponents $X(N)$ in a very simple manner:
\begin{equation}
	\overline{C^N(u,u+R)}\equiv\int dC C^N P_R(C)\sim 
e^{-\frac{2\pi}LX(N)R},
\label{eqCstrip}
\end{equation}
In the strip geometry, the system thus exhibits an even closer relation to
the dynamical model of section~\ref{subsec:NonEq}, with the role of time $t$
now played by the coordinate along the strip.
In figure \ref{fig3}, we have plotted $\Delta(N)$ obtained from the exponents
$X(N)$ using the definition Eq. \ref{deltaetX}. The original data are plotted
as filled black symbols while the exponents $\Delta(N)$ after the 
transformation
$1-N\rightarrow N$ are in open red  symbols. 
Within numerical accuracy, the symbols fall
on top of each other as expected from Eq. \ref{symdeltaN}. 

\begin{figure}[!th]
\epsfxsize=16cm\epsfbox{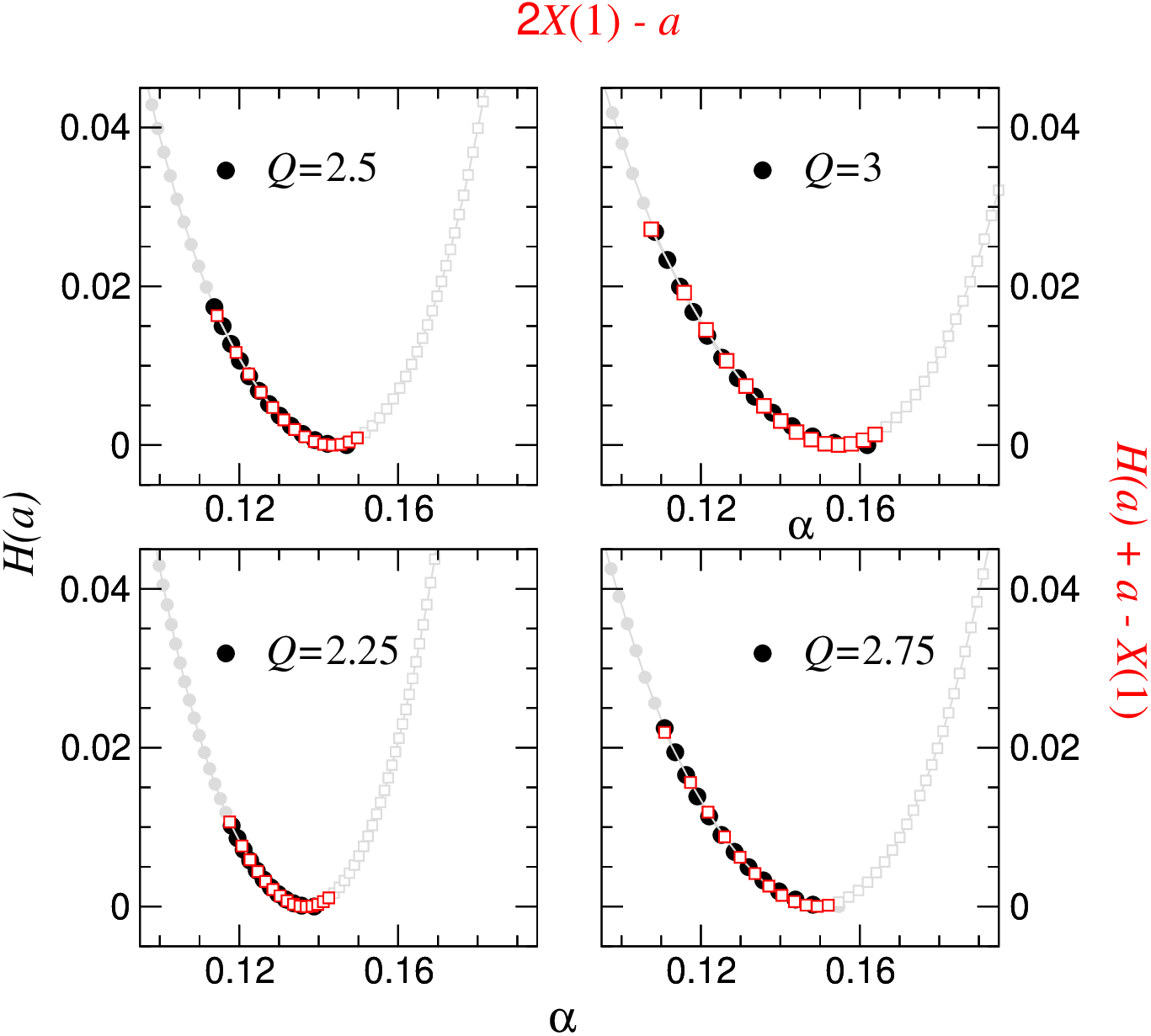}
\caption{Numerical test of the spectrum symmetry as calculated from
the set of exponents $X(N)$ obtained from the decay of the moments
$\overline{\langle \sigma_i\sigma_{i+u}\rangle^N}$ for the self-dual
random bond Potts model for $Q\le3$. The filled circles correspond to
$H(a)$ vs $a$, while the open squares correspond to $H(a)+a-X(1)$
vs $2X(1)-a$. The part of the curves which superimpose (black and
red online) is emphasized.}
\label{fig1}
\end{figure}

The spectral function $H(a)$ follows from the expression Eq. \ref{eqCstrip} of the
spin-spin correlation function on the strip:
\begin{equation}
H(a)=X(N)-aN.
\label{eqH}
\end{equation}
We present on figure \ref{fig1} the multifractal spectrum $H(a)$ for the
self-dual random-bond Potts model. This spectrum has been obtained
using Eq. \ref{eqH} and the exponents $X(N)$ computed by interpolation
of the moments of the spin-spin correlation function with Eq. \ref{eqCstrip}.
On top of the multifractal spectrum $H(a)$, full symbols on the figure,
we have plotted as open symbols the image of $H(a)$ under the transformations
$H\left(a\right)+a-X(1)\rightarrow H(a)$ and $2X(1)-a\rightarrow a$.
If the symmetry holds for the random-bond Potts model, the two curves
should fall on top of each others according to Eq. \ref{symHapotts}.
We have emphasized with black and red colors on the figure the region
where the two curves superimpose. The numerical data are in very
good agreement with Eq. \ref{symHapotts} for $Q$ from $2.25$ to $3$.

\begin{figure}[!th]
\epsfxsize=14cm\epsfbox{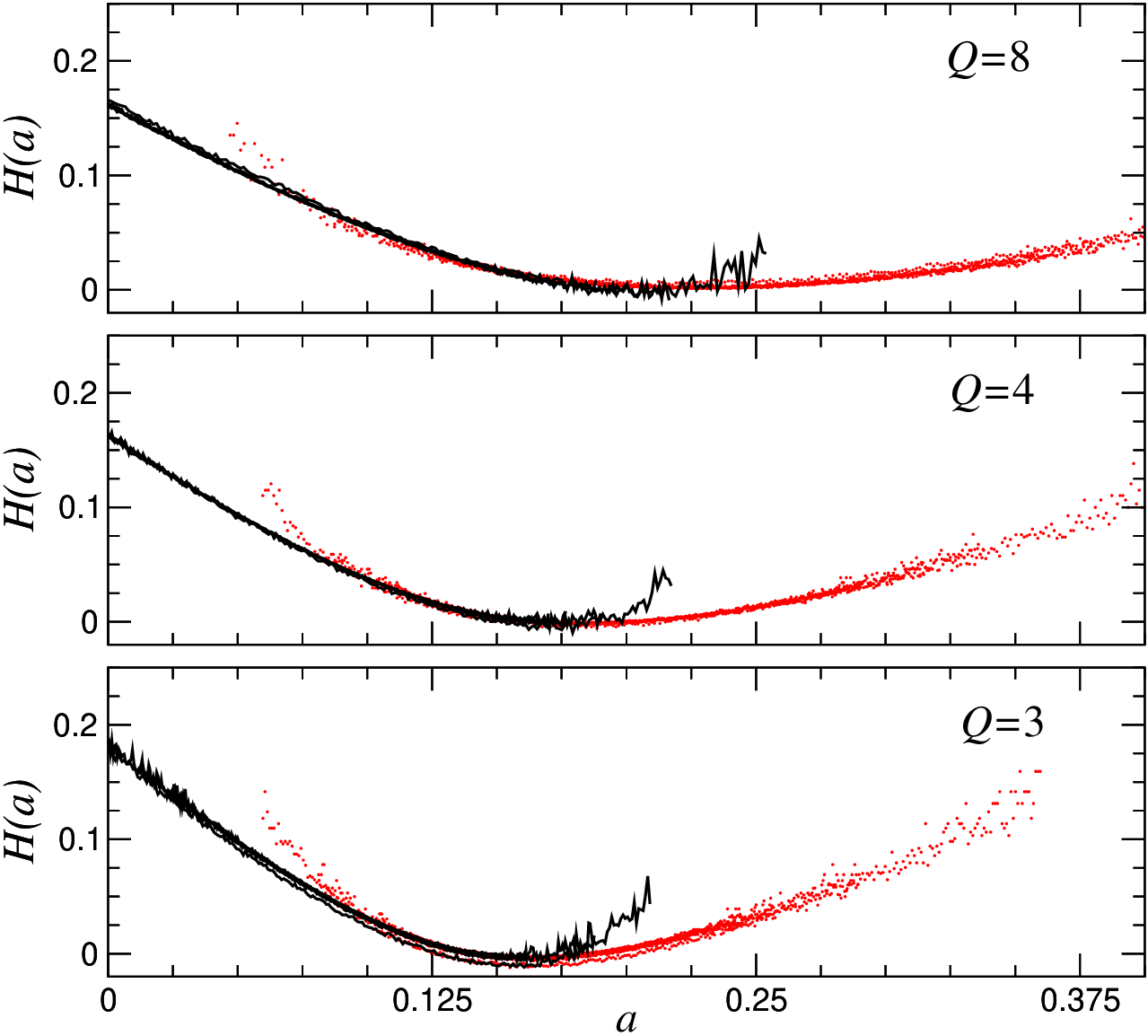}
\caption{Numerical test of the spectrum symmetry as calculated from
the probability distribution of the correlation function of the dilute
bond Potts model for $Q\ge 3$. The full lines (black online) 
correspond to $H(a)$ vs $a$, while the filled circles (red online)
correspond to $H(a)+a-X(1)$ vs $2X(1)-a$. } 
\label{fig2}
\end{figure}

We now present another test of the symmetry using an alternative determination
of the multifractal spectrum $H(a)$.  The definition \ref{eqCstrip}, written
in terms of the variable $Y=-\ln C$ reads as a Laplace transform
\begin{equation}
\int dY P_R(Y)e^{-NY}\sim e^{-\frac{2\pi}LX(N)R},
\end{equation}
The inverse Laplace transform evaluated close to the saddle point
approximation leads to (see Ref.~\cite{Cha_Ber_Sh})
\begin{equation}
P_R(a)\sim\left(\frac{2\pi R}{L}\right)^{1/2}e^{-\frac{2\pi R}{L}H(a)}
\label{eqPrStrip}
\end{equation}
with $a=-\ln C/(2\pi R/L)$.
The spectral function is thus directly given in terms of the probability
distribution $P_R(-\ln C/(2\pi R/L))$. Using again the connectivity 
transfer matrix,
we have computed the probability distribution of the spin-spin correlation
function for the diluted Potts model~\cite{Cha_Ber_Sh}.
The multifractal spectrum was then obtained by interpolation of the numerical data with
Eq. \ref{eqPrStrip}. In figure \ref{fig2}, the multifractal spectrum is
plotted in black for three different values of the number of states $Q$.
In red, we have plotted the data after the transformations
$H\left(a\right)+a-X(1)\rightarrow H(a)$ and $2X(1)-a\rightarrow a$.
Again, the data fall nicely on top of each other for small and intermediate
values of $a$. For large values of $a$, the two curves significantly
deviate from each other, but at the same time, the data dispersion which
measures the uncertainty also increases. 
According to \ref{eqH}, this region, where the
multifractal spectrum increases with $a$, is associated to negative moments
$N$ of the spin-spin correlation function. The latter are dominated by rare
events corresponding to weak spin-spin correlation and thus strongly diluted
realisations of disorder. Numerically, one can hope to sample correctly events
whose probability is of order $1/N_{\rm Dis.}$ where $N_{\rm Dis.}$ is
the number of disorder realizations considered.
Very rare events are unlikely to be generated. We may thus interpret the
deviation observed on figure \ref{fig2} as due to an underestimate of the
tail of the probability distribution $P(C) \lesssim 10^{-5}$.

\subsection{Reminder on existing perturbative expansions in $(Q-2)$}

The Harris criterion indicates that disorder is relevant at the transition
for $Q>2$ but only marginal for $Q=2$: the point $Q=2$ can be thus used
as a starting point for a perturbative expansion in 
the parameter (which measures the relevance of disorder)
	\begin{equation}
y_H=\alpha/\nu=\frac{4}{3\pi}(Q-2)-\frac{4}{9 \pi^2} (Q-2)^2 +O(Q-2)^3	
	\label{yH}
	\end{equation}
which measures the disorder relevance \cite{Ludwig}.

\subsubsection{ First order contribution }

The exponents $X(N)$ governing the correlation functions 
have been computed at first order in $y_H$ by Ludwig~\cite{Ludwig}:
	\begin{equation}
	X(N)=NX(1)-\frac{N(N-1)}{ 16}y_H+{\cal O}(y_H^2)
	\label{ExpLudwig}
	\end{equation}
The corresponding Gaussian 
multifractal spectrum $H(\alpha)$ can be found in
Ludwig~\cite{Ludwig}, but the
symmetry is much easily tested under the form of Eq. \ref{symdeltaN}.
According to the definition Eq. \ref{deltaetX}, 
the expansion Eq. \ref{ExpLudwig} leads to
	\begin{equation}
	\Delta(N)=X(N)-NX(1)=-\frac{N(N-1)}{ 16}y_H+{\cal O}(y_H^2)
	\end{equation}
which is manifestly invariant under the transformation $N\rightarrow 1-N$.

\subsubsection{Second order contribution  }

 The perturbative 
expansion of Ludwig has been extended to the second order in a replica 
symmetric scenario
by Lewis~\cite{Lewis}:
	\begin{equation}
	X(N)=NX(1)-\frac{N(N-1)}{ 16}\left[y_H+\left(\frac{11}{ 12}-4\ln 2
	+\frac{N-2}{ 24}\left(33-\frac{29\pi}{\sqrt 3}\right)\right)
         \frac{y_H^2}{ 2}\right]
	+{\cal O}(y_H^3)
	\label{resLewis}
	\end{equation}
The presence of the $(N-2)$ term explicitly breaks 
the symmetry $N\to 1-N$.
 As a comparison, the equivalent weak-multifractality
perturbative expansion for Anderson localization
can be found in \cite{Wegner_epsilon} up to order $\epsilon^4$
 in $\epsilon=d-2$ and satisfy the multifractal symmetry.

If the result of Eq. \ref{resLewis} is really true, this means
that the multifractal symmetry is not realized in the random Potts model.
However, we believe that one should be cautious concerning 
the result of Eq. \ref{resLewis}, in spite of previous numerical 
confirmations. 
It is a
 perturbative replica result that has been derived within the
 so-called replica-symmetric scenario.
Indeed in some disordered systems, the replica symmetry
 is known to be spontaneously broken, so that one should in principle
study the various fixed points corresponding
to various replica symmetry properties
and then decide which one corresponds to the physical fixed point.
For the random Potts model, two different types of fixed 
points have been compared in~\cite{DotsenkoEtAl}:
 the replica symmetric fixed point, used to derive Eq. \ref{resLewis}
and the broken replica symmetry fixed point based
on the Parisi block diagonal matrix Ansatz.
We refer to \cite{Dot_Har} for a detailed discussion of the
possible physical meanings of replica symmetry breaking
at the critical point of random ferromagnets.
According to ~\cite{DotsenkoEtAl}, the differences 
in the critical exponents between these two fixed points
appear only at second order $y_H^2$, so that 
the first order contribution of Eq. \ref{ExpLudwig}, although derived
within the replica symmetric fixed point, seems to be of larger validity
in the space of replica possible fixed points,
whereas the second order contribution of Eq. \ref{resLewis} 
seems specific to the replica symmetric fixed point.
Since it has not been possible up to now to determine
on theoretical grounds
which fixed point describes the scaling limit of
the disordered lattice models of Eq. \ref{defPotts}, very precise
numerical computations ~\cite{Cha_Ber,Jacobsen}
have been necessary to discriminate
between the two fixed points studied in ~\cite{DotsenkoEtAl}:
these numerical studies ~\cite{Cha_Ber,Jacobsen} have concluded
(i) that the replica symmetric broken fixed point based
on the Parisi block diagonal matrix Ansatz is excluded,
and (ii) that the replica symmetric fixed point is thus the physical one.
However we believe that (i) does not directly imply (ii), because
only two particular fixed points have been studied. Since the 
Parisi block diagonal matrix Ansatz comes from spin-glasses
in the mean-field limit, whose physics is completely different
from two-dimensional random ferromagnets, one may argue that
another type of replica symmetry breaking
could be realized in critical random ferromagnets.
For instance, it has been proposed in \cite{DotMez}
that in some disordered systems, it was appropriate to consider
some vector symmetry breaking scheme instead of the usual Parisi
matrix sector breaking Ansatz. Very recently, still another pattern
of replica symmetry breaking has been identified 
to account for a pre-freezing phenomenon of multifractal exponents
in a disordered system \cite{fyodorov}. As a consequence,
the proper identification of the true physical fixed point
among the possible fixed points in replica space 
remains the important issue to decide whether Eq. \ref{resLewis}
is really true and sufficient to rule out
 the multifractal symmetry.

\subsection{ Discussion}

In summary, we cannot give a definite conclusion
on the existence of the multifractal symmetry in 
the random Potts model,
but we would like to stress the two following points:

(1) from a purely numerical point of view, the data presented
on the Figures above satisfy the symmetry within the error bars
for all values $2 < Q \leq 8$ that have been studied. This statement is
valid for the self dual disordered model and in the dilute case.
 Moreover, the differences between
the numerical data and the second-order perturbative expansion
of Eq. \ref{resLewis} plotted as 
dashed lines in figure~\ref{fig3}, 
are significantly larger than the deviation
from a symmetric $\Delta(N)$.

(2) from a theoretical point of view, it would be interesting
to study other fixed points that break the replica symmetry
using various schemes, to see whether one can identify a type of scheme
that would preserve the multifractal symmetry.

\section{ Conclusions and perspectives }
\label{secconclusion}

In this paper, we have analyzed the physical origin
of the symmetry relation $f(2d-\alpha)=f(\alpha)+d-\alpha$
proposed by Mirlin, Fyodorov, Mildenberger and Evers \cite{mirlin06}
for the singularity spectrum $f(\alpha)$ of  critical eigenfunctions at Anderson transitions.
We have explained the analogy with the Gallavotti-Cohen symmetry 
of large deviation
functions that are well-known in the field of non-equilibrium dynamics: 
the multifractal spectrum of the disordered model
corresponds to the large deviation function of the
rescaling exponent $\gamma=(\alpha-d)$ for a renormalization procedure
considered as a Markov chain in the 'effective time' 
$t=\ln L$.  We have concluded that 
the symmetry discovered in the specific example of Anderson transitions
should actually be satisfied at many other random critical
points after an appropriate translation.
For many-body random phase transitions,
where the critical properties are usually analyzed in terms of
 the multifractal spectrum $H(a)$ and of the moments exponents $X(N)$
of two-point correlation function, we have obtained that the symmetry becomes
 $H( 2X_1 -a)= H( a )  + a-X(1)$,
or equivalently $\Delta(N)=\Delta(1-N)$ for the anomalous parts
$\Delta(N) \equiv  X(N)-NX(1)$. 
We have presented detailed numerical tests in favor
of these relations, or at least compatible with them, 
for the two-dimensional random Potts model
 with various $Q>2$.
Although presently available numerical results for the scaling dimensions 
are in favour of the RG results
deduced from the replica symmetric scenario (which does not satisfy the
multifractal symmetry) and rule out the Parisi matrix sector breaking Ansatz, 
we cannot exclude that other
fixed points would equally be compatible with the numerical results, but also 
with an exact multifractal symmetry.
It would be interesting in the future to test this multifractal 
 symmetry at other random critical points.
 At a more general level, we hope 
that the analogy with the field of non-equilibrium dynamics
 will give new insights
 into the properties of renormalization flows in disordered systems.

\end{document}